\documentclass[twoside,twocolumn,9pt]{article}
\usepackage{extsizes}
\usepackage[super,sort&compress,comma]{natbib} 
\usepackage[version=3]{mhchem}
\usepackage[left=1.5cm, right=1.5cm, top=1.785cm, bottom=2.0cm]{geometry}
\usepackage{balance}
\usepackage{mathptmx}
\usepackage{sectsty}
\usepackage{graphicx} 
\usepackage{lastpage}
\usepackage[format=plain,justification=justified,singlelinecheck=false,font={stretch=1.125,small,sf},labelfont=bf,labelsep=space]{caption}
\usepackage{float}
\usepackage{fancyhdr}
\usepackage{fnpos}
\usepackage[english]{babel}
\addto{\captionsenglish}{%
  
}
\usepackage{array}
\usepackage{droidsans}
\usepackage{charter}
\usepackage[T1]{fontenc}
\usepackage[usenames,dvipsnames]{xcolor}
\usepackage{setspace}
\usepackage[compact]{titlesec}
\usepackage{hyperref}
\usepackage{comment}

\usepackage{epstopdf}%This line makes .eps figures into .pdf - please comment out if not required.

\definecolor{cream}{RGB}{222,217,201}

\begin{document}

\pagestyle{fancy}
\thispagestyle{plain}
\fancypagestyle{plain}{
%%%HEADER%%%
\renewcommand{\headrulewidth}{0pt}
}
%%%END OF HEADER%%%

%%%PAGE SETUP - Please do not change any commands within this section%%%
\makeFNbottom
\makeatletter
\renewcommand\LARGE{\@setfontsize\LARGE{15pt}{17}}
\renewcommand\Large{\@setfontsize\Large{12pt}{14}}
\renewcommand\large{\@setfontsize\large{10pt}{12}}
\renewcommand\footnotesize{\@setfontsize\footnotesize{7pt}{10}}
\makeatother

\renewcommand{\thefootnote}{\fnsymbol{footnote}}
\renewcommand\footnoterule{\vspace*{1pt}% 
\color{cream}\hrule width 3.5in height 0.4pt \color{black}\vspace*{5pt}} 
\setcounter{secnumdepth}{5}

\makeatletter 
\renewcommand\@biblabel[1]{#1}            
\renewcommand\@makefntext[1]% 
{\noindent\makebox[0pt][r]{\@thefnmark\,}#1}
\makeatother 
\renewcommand{\figurename}{\small{Fig.}~}
\sectionfont{\sffamily\Large}
\subsectionfont{\normalsize}
\subsubsectionfont{\bf}
\setstretch{1.125} %In particular, please do not alter this line.
\setlength{\skip\footins}{0.8cm}
\setlength{\footnotesep}{0.25cm}
\setlength{\jot}{10pt}
\titlespacing*{\section}{0pt}{4pt}{4pt}
\titlespacing*{\subsection}{0pt}{15pt}{1pt}
%%%END OF PAGE SETUP%%%

%%%FOOTER%%%
\fancyfoot{}
%\fancyfoot[LO,RE]{\vspace{-7.1pt}\includegraphics[height=9pt]{head_foot/LF}}
%\fancyfoot[CO]{\vspace{-7.1pt}\hspace{11.9cm}\includegraphics{head_foot/RF}}
%\fancyfoot[CE]{\vspace{-7.2pt}\hspace{-13.2cm}\includegraphics{head_foot/RF}}
\fancyfoot[RO]{\footnotesize{\sffamily{1--\pageref{LastPage} ~\textbar  \hspace{2pt}\thepage}}}
\fancyfoot[LE]{\footnotesize{\sffamily{\thepage~\textbar\hspace{4.65cm} 1--\pageref{LastPage}}}}
\fancyhead{}
\renewcommand{\headrulewidth}{0pt} 
\renewcommand{\footrulewidth}{0pt}
\setlength{\arrayrulewidth}{1pt}
\setlength{\columnsep}{6.5mm}
\setlength\bibsep{1pt}
%%%END OF FOOTER%%%

%%%FIGURE SETUP - please do not change any commands within this section%%%
\makeatletter 
\newlength{\figrulesep} 
\setlength{\figrulesep}{0.5\textfloatsep} 

\newcommand{\topfigrule}{\vspace*{-1pt}% 
\noindent{\color{cream}\rule[-\figrulesep]{\columnwidth}{1.5pt}} }

\newcommand{\botfigrule}{\vspace*{-2pt}% 
\noindent{\color{cream}\rule[\figrulesep]{\columnwidth}{1.5pt}} }

\newcommand{\dblfigrule}{\vspace*{-1pt}% 
\noindent{\color{cream}\rule[-\figrulesep]{\textwidth}{1.5pt}} }

\makeatother
%%%END OF FIGURE SETUP%%%

%%%TITLE, AUTHORS AND ABSTRACT%%%
\twocolumn[
  \begin{@twocolumnfalse}
%{\includegraphics[height=30pt]{head_foot/PCCP}\hfill\raisebox{0pt}[0pt][0pt]{\includegraphics[height=55pt]{head_foot/RSC_LOGO_CMYK}}\\[1ex]
%\includegraphics[width=18.5cm]{head_foot/header_bar}}\par
\vspace{1em}
\sffamily
\begin{tabular}{m{4.5cm} p{13.5cm} }

 & \noindent\LARGE{\textbf{Probing ultracold chemistry using ion spectrometry}} \\%Article title goes here instead of the text "This is the title"
\vspace{0.3cm} & \vspace{0.3cm} \\

 & \noindent\large{Yu Liu,$^{\ast}$\textit{$^{b}$}\textit{$^{a}$}\textit{$^{c}$} David D. Grimes,$^{\ast}$\textit{$^{a}$}\textit{$^{b}$}\textit{$^{c}$} Ming-Guang Hu$^{\ast}$\textit{$^{a}$}\textit{$^{b}$}\textit{$^{c}$} and Kang-Kuen Ni\textit{$^{a}$}\textit{$^{b}$}\textit{$^{c}$}} \\%Author names go here instead of "Full name", etc.

 & \noindent\normalsize{Rapid progress in atomic, molecular, and optical (AMO) physics techniques enabled the creation of ultracold samples of molecular species and opened opportunities to explore chemistry in the ultralow temperature regime. In particular, both the external and internal quantum degrees of freedom of the reactant atoms and molecules are controlled, allowing studies that explored the role of the long-range potential in ultracold reactions. The kinetics of these reactions have typically been determined using the loss of reactants as proxies. To extend such studies into the short-range, we developed an experimental apparatus that combines the production of quantum-state-selected ultracold KRb molecules with ion mass and kinetic energy spectrometry, and directly observed KRb + KRb reaction intermediates and products [Science, 2019, \textbf{366}, 1111]. Here, we present the apparatus in detail. For future studies that aim for detecting the quantum states of the reaction products, we demonstrate a photodissociation based scheme to calibrate the ion kinetic energy spectrometer at low energies.} \\%The abstract goes here instead of the text "The abstract should be..."

\end{tabular}

 \end{@twocolumnfalse} \vspace{0.6cm}

  ]
%%%END OF TITLE, AUTHORS AND ABSTRACT%%%

%%%FONT SETUP - please do not change any commands within this section
\renewcommand*\rmdefault{bch}\normalfont\upshape
\rmfamily
\section*{}
\vspace{-1cm}

%%%FOOTNOTES%%%

\footnotetext{\textit{$^{a}$~Department of Chemistry and Chemical Biology, Harvard University, Cambridge, Massachusetts 02138}}
\footnotetext{\textit{$^{b}$~Department of Physics, Harvard University, Cambridge, Massachusetts 02138}}
\footnotetext{\textit{$^{c}$~Harvard-MIT Center for Ultracold Atoms, Cambridge, Massachusetts 02138}}
\footnotetext{\textit{$^{\ast}$}~These authors contributed equally to this work}

%%%END OF FOOTNOTES%%%

%%%MAIN TEXT%%%%

\section{Introduction} \label{section:intro}

Over the past decades, colder and more precisely quantum-controlled molecular samples have been hotly pursued in the AMO community for a diverse range of applications including precision measurements \cite{baron2014order,cairncross2017precision,andreev2018improved}, quantum simulations \cite{micheli2006toolbox,buchler2007strongly, cooper2009stable}, and quantum computation \cite{demille2002quantum,yelin2006schemes,ni2018dipolar,hudson2018dipolar}. These works aim to take advantage of the electric dipole moments and manifolds of internal states possessed by molecules as reviewed in a previous PCCP perspective~\cite{ni2009dipolar} and other references~\cite{carr2009cold,lique2017cold}. Furthermore, cold molecules offer a new platform to explore chemistry \cite{balakrishnan2016perspective}, which is the main topic of this perspective. While the ``ultracold'' regime is characterized by single partial wave collisions ($s$-wave for identical bosons and distinguishable particles, $p$-wave for identical fermions), which typically occurs below 1 millikelvin, our discussion here also  includes the ``cold'' regime, which loosely refers to collision energies up to a few kelvin.

\begin{figure*}
\centering
\includegraphics[width=6.5 in]{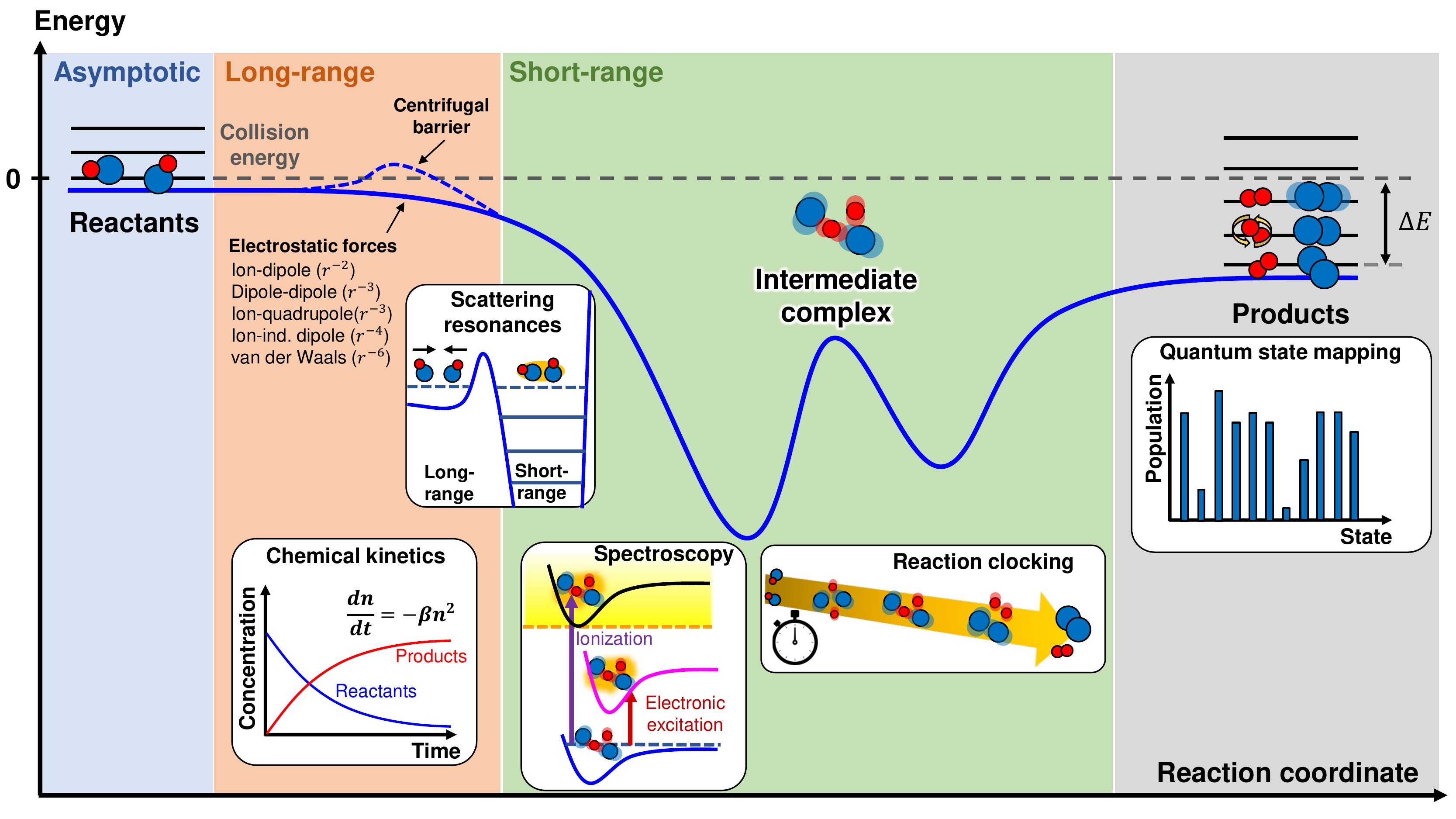}
\caption{Schematic PES (not to scale) of a barrierless, exothermic cold reaction along some reaction coordinate, divided into asymptotic, long-range, and short-range portions. Experimental techniques that are sensitive to the different portions are illustrated in white boxes accordingly. Reactants are prepared with low collision energies and in well-defined quantum states in the asymptotic region. They approach each other under the influence of the long-range potential, characterized by centrifugal and electrostatic forces. The potential well in the short-range leads to the formation of a transient intermediate complex, which will proceed further to products in the case of exothermic reactions, or back to reactants in endothermic ones. Though cold reactions are overall barrierless, submerged barriers, such as illustrated above, are predicted to exist in many reactions \cite{tscherbul2008formation, byrd2010structure,byrd2012structure,chen2019isotope}.}
\label{figPES}
\end{figure*}

To understand chemical reactions in the cold regime, we divide the underlying potential energy surface (PES) into asymptotic, long-range, and short-range portions~\cite{quemener2012ultracold} as shown in Fig~\ref{figPES}. Reactants are prepared in well-defined quantum states in the asymptotpic region and are set on a collision course. Long-range forces, such as centrifugal and electrostatic terms, govern the approach between the reactants, while short-range forces determine the dynamics of the intermediate complex and the formation of the products. Since the collision energy is extremely low, a reaction can occur only if the short-range is a potential well, not a barrier. As a result, reactants that proceed into the short-range will form a transient intermediate complex. This complex, in the absence of dissipative processes, will either dissociate back into reactants (for endothermic reactions) or continue to form products (for exothermic reactions). Below, we survey past and ongoing work for studying long- and short-range dynamics in cold molecular systems, with relevant concepts and techniques illustrated in Fig. \ref{figPES}.

At sufficiently low temperatures, the collision energy is comparable or smaller than the energy scale of the long-range interaction between reactants. As a result, long-range forces can profoundly influence the rates and outcomes of chemical reactions, unlike in higher temperature reactions. Moreover, the form of the long-range PES is in many cases directly tied to the internal quantum states of the reactants, enabling quantum control over reactions through internal state manipulation. For example, quantum statistics of the reactants can dictate the lowest partial waves allowed in collisions. For $s$-wave, the reactants can proceed directly to the short-range, while for $p$-wave they must tunnel through a centrifugal barrier to reach the short-range. Consequently, quantum statistics controlled reaction rates that differ by a factor of 10-100 were observed~\cite{ospelkaus2010quantum}. Introducing dipole-dipole interactions by polarizing the molecules in an electric field can significantly enhance their chemical reactivity \cite{ni2010dipolar} and additionally confining them in reduced dimensions can further alter their reactivity \cite{de2011controlling}. Large dipolar interactions can lead to collisions with contributions from many partial waves, resulting in highly anisotropic interactions between molecules \cite{guo2018dipolar}. Long-range interactions can be further modified by electronically exciting one of the reactants, in which case effects like the radiative lifetime \cite{puri2019reaction} and orbital shape \cite{hall2012millikelvin} of the excited state become significant to subsequent reaction rates.

In all of the above studies, reactant loss or product appearance serve as proxies to the overall reaction rate determined by the long-range potential. In the case of scattering resonances, the reaction rate also depends sensitively on the short-range PES. A resonance arises when the scattering state of the reactants or products becomes strongly coupled to a bound or quasi-bound state of the intermediate complex. Resonances involving quasi-bound states are known as ``shape resonances'', while those involving true bound states are known as ``Feshbach resonances''. While resonances are typically obscured by the effects of thermal averaging in higher temperature reactions, they become resolved in cold reactions where collision energies are low and narrowly distributed. Observations of shape resonances have been made in both photodissociation \cite{mcdonald2016photodissociation} and Penning ionization \cite{klein2017directly} reactions, providing benchmark tests for the accuracy of the highest-level $\textit{ab initio}$ surfaces. Feshbach resonances play an important role in ultracold collisions of atoms~\cite{chin2010feshbach}, and have been predicted for both atom-molecule \cite{mayle2012statistical,hummon2011cold,frye2016approach} and molecule-molecule \cite{bohn2002rotational,tscherbul2009magnetic} collisions. Magnetically tunable Feshbach resonances have been observed in ultracold NaK + K collisions \cite{yang2019observation}, probing the short-range PES with exceptional resolution and challenging quantum chemistry calculations for heavy atom systems.

Beyond studying the chemical kinetics of reactants and products, more can be learned about the short-range dynamics directly by studying the intermediate complex, which is challenging due to its transient nature. The lifetime of the complex, $\tau_c$, is proportional to the density, $\rho_c$, of internal modes through which the energy of the complex can be re-distributed, and inversely proportional to number of exit channels, $N_0$, available for the complex to dissociate, either into products or reactants \cite{levine2009molecular}. For reactions of small systems, $\tau_c$ is typically no more than on the order of a rotational period (10 ps), and a direct observation of the complex requires ultrafast techniques \cite{scherer1990real}. For larger systems the relatively large number of internal modes can extend the lifetime to as long as microseconds \cite{noll1998bimolecular}. In the ultracold regime, the lifetime of the complex becomes substantial for even small systems by preparing the the reactants in their absolute ground rovibronic state and thereby minimizing $N_0$. This enabled our direct observation of the intermediate complex K$_2$Rb$_2^\ast$, formed by the bimolecular reaction between KRb molecules, without the use of ultrafast lasers \cite{hu2019direct}. Through measurement of the equilibrium concentration of the complex, we estimated the lifetime of the complex to be on the order of a few hundred nanoseconds to a few microseconds. The observation of the complex opens up the exciting possibility to directly probing the short-range of the reaction. In the time-domain, we can ``clock'' the reaction and make a direct and more precise measurement of complex lifetime. In the frequency-domain, we can learn more about the structure of the complex through spectroscopy.

The short-range PES not only determines the dyanmics of the complex, but also guides the process of product formation. As such, studying the quantum states of the products can provide additional information about the short-range. Product state mapping, while enjoying a long and successful history as a workhorse technique behind many studies of reaction dynamics in physical chemistry at large\cite{ashfold2006imaging}, has seen very few applications to cold chemistry. ``The detection of product states'' was recognized as one of the ``major milestones that need to be reached for the continued progress of this field'' by a PCCP editorial on cold molecules in 2011 \cite{dulieu2011physics}. A challenge to reaching this milestone in AMO experiments is that the techniques used to create and control the reactants are optimized for probing reaction species in a quantum state-specific manner and do not have the flexibility to study more than a handful of quantum states at a time, whereas typical chemical reactions liberate enough energy to populate hundreds of product quantum states or more. Reactions involving single \cite{rui2017controlled} \cite{hoffmann2018reaction} or relatively few \cite{wolf2017state} channels are still amenable to product-state specific studies using AMO techniques. More generally, however, mapping out the product state distribution demands an integration of physical chemistry detection techniques into the AMO experimental infrastructure.

Ultimately, investigating both the intermediate and products will provide the most comprehensive picture on the short-range dynamics. This is what we hope to achieve with the ``ultracold chemistry machine'' to be described in detail in section \ref{section:combine AMO & PChem}.

\section{The platform: KRb, bialkalis, and beyond} \label{section:case for KRb}

We began our exploration of ultracold chemistry by studying the exchange reaction between ground state potassium-rubidium molecules

\begin{equation} \label{KRb+KRb}
^{40}\textrm{K}^{87}\textrm{Rb} (v = 0, j = 0) + ^{40}\textrm{K}^{87}\textrm{Rb} (v = 0, j = 0) \rightarrow \textrm{K}_{2}\textrm{Rb}_{2}^* \rightarrow \textrm{K}_2 + \textrm{Rb}_2,
\end{equation}

\noindent where $ v $ and $ j $ denote the rotational and vibrational quantum numbers of the reactants, respectively. The reaction is exothermic with $ \Delta E = - 10.4(4) ~\textrm{cm}^{-1}$.\cite{ospelkaus2010quantum} This reaction was first inferred by Ospelkaus \textit{et al.} in 2010 through the detection of reactant (KRb) loss\cite{ospelkaus2010quantum}, and recently confirmed by the detection of the reaction products (K$_2$ and Rb$_2$) and the transient intermediate complex (K$_2$Rb$_2^*$) in our recent work \cite{hu2019direct}. The short-range dynamics of KRb + KRb has been the subject of a number of theoretical studies. Mayle \textit{et al.}\cite{mayle2013scattering} suggests that the long-lived intermediate complex (K$_2$Rb$_2^*$) will ergodically explore the available reaction phase space and the dynamics can be adequately captured by statistical theory. In this case the complex lifetime can be simply calculated using the RRKM theory as $\tau_c = h \rho_c / N_0$, where $h$ is Planck's constant. Based on the assumption of statistical behavior, Gonzalez \textit{et al.}\cite{gonzalez2014statistical} calculated the quantum state distribution of the products. Experimentally, the low exothermicity, and therefore the few accessible product exit channels, makes the reaction amenable to full product quantum state mapping. Measuring both the product state distribution and the complex lifetime will provide fundamental tests for the applicability of statistical theory to this reaction. Detecting departures from statistical behavior would in itself be interesting, and reasons for expecting such departures are suggested by Nesbitt\cite{nesbitt2012toward}.

In addition to the molecule-molecule reaction, we can also investigate the atom-molecule reactions

\begin{align} \label{K+KRb}
    \textrm{K} + \textrm{KRb} & \rightarrow \textrm{K}_2\textrm{Rb}^* \rightarrow \textrm{K}_2 + \textrm{Rb},
\end{align}

and

\begin{align} \label{Rb+KRb}
    \textrm{Rb} + \textrm{KRb} & \rightarrow  \textrm{K}\textrm{Rb}_2^* \rightarrow \textrm{Rb}_2 + \textrm{K}.
\end{align}

\noindent For ground-state reactants, reaction \ref{K+KRb} is exothermic with $\Delta E = -224.972(4) ~\textrm{cm}^{-1}$,\cite{ospelkaus2010quantum} and reaction \ref{Rb+KRb} is endothermic with $\Delta E = 214.6(4) ~\textrm{cm}^{-1}$.\cite{ospelkaus2010quantum} Reaction \ref{K+KRb} was first identified again by Ospelkaus \textit{et al.} through the observation of rapid KRb loss in the presence of K atoms \cite{ospelkaus2010quantum}. While a four-atom reaction involving heavy alkali atoms such as reaction \ref{KRb+KRb} is still beyond the current scope of quantum chemistry calculations, numerically-exact quantum dynamics calculations have been reported for K + KRb \cite{croft2017universality}. Product state mapping of reaction \ref{K+KRb} will provide valuable benchmark to the state-of-the-art theory.

KRb is a member of the bialkali family of molecules, which are among the first molecular species to be brought into the ultracold regime \cite{nikolov1999observation,wang2004photoassociative,mancini2004observation,deiglmayr2008formation} and, to date, remain the densest, coldest, and among the best quantum state controlled molecular samples \cite{ospelkaus2010quantum,danzl2010ultracold,park2015ultracold,guo2016creation}. They are a major platform in ultracold chemistry research, including most of the studies mentioned in the introduction (section \ref{section:intro}). While bialkalis will continue to serve as a rich playground for chemical dynamics for years to come, efforts are well underway to bring more ``chemically relevant'' species into the fold. The trapping of O$_2$ \cite{akerman2017trapping} in a superconducting magnetic trap and the observation of O$_2$ - O$_2$ collisions \cite{segev2019collisions} represent a major milestone in increasing the chemical diversity of ultracold chemistry.

\section{The ultracold chemistry machine} \label{section:combine AMO & PChem}

\begin{figure*}
\centering
\includegraphics[width=6.5 in]{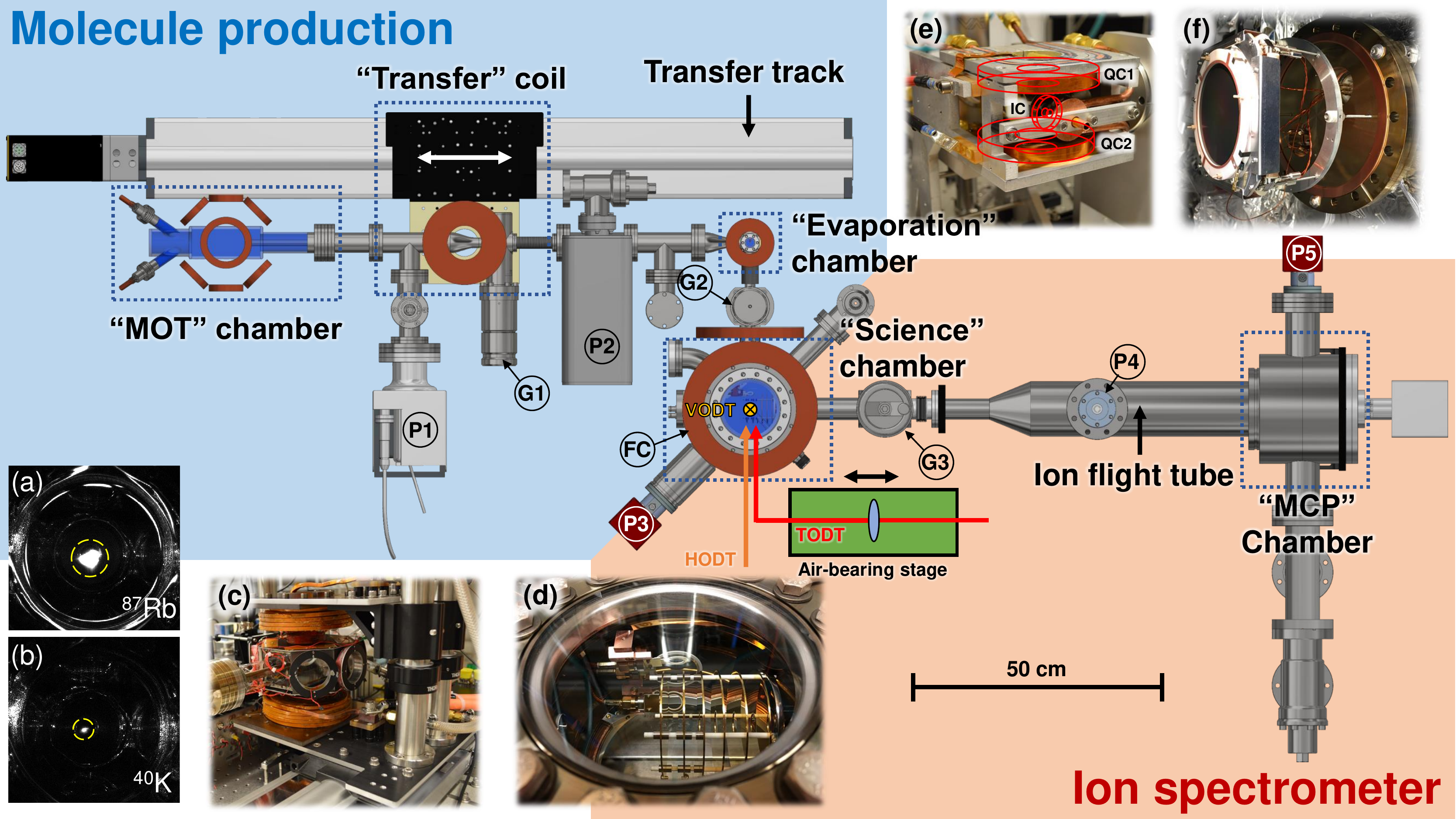}
    \caption{The ultracold chemistry machine. A bird's-eye view of the apparatus is shown, with key components highlighted and labeled. Supporting structures for the vacuum chambers and magnetic coils are hidden in the interest of clarity. G1-3: gate valves; P1-5: vacuum ion pumps; FC: Feshbach magnetic field coil pair. (a) (b) Fluorescence images of a $^{87}$Rb MOT with $ \sim 1\times 10^9$ atoms and $^{40}$K MOT with $ \sim 5\times 10^6$ atoms in the ``MOT'' chamber. (c) Quadrupole magnetic ``transfer'' coil pair. (d) VMI ion optics viewed though the top viewport of the ``science'' chamber. (e) QUIC magnetic trap surrounding the evaporation chamber. the trap consists of three coils labeled as QC(quadrupole coil)1, QC2, and IC(Ioffe coil). red wireframes indicate the locations of the coils in the picture. (f) The MCP detector assembly that is housed inside the ``MCP'' chamber.}
\label{figMachine}
\end{figure*}

As we have illustrated in the introduction (section \ref{section:intro}), part of what makes exploring chemistry in the ultracold regime exciting is the ability to start a reaction with the reactants prepared in well-defined quantum states. This is made possible by the high degree of quantum state control in AMO experiments. Ultracold molecules are typically detected in these experiments through optical (fluorescence and absorption) imaging \cite{ketterle1999making}. This technique provides powerful visualizations of the spatial configurations of particles and information such as their number, density, momentum, and temperature. A key requirement for imaging ultracold matter, where the typical density ranges from $10^{8}$ to 10$^{14}$~cm$^{-3}$, is that each molecule can scatter many photons to provide sufficient signal for detection. This demands molecules prepared in a specific (oftentimes the ground rovibronic) quantum state to repeatedly undergo ``cycling'' optical transition. While such transitions are found in many atoms, they are rare in molecules \cite{shuman2010laser, zhelyazkova2014laser,anderegg2017radio,kozyryev2017sisyphus}. This requirement can be side-stepped in ultracold bialkali molecules, such as KRb, as they can be coherently dissociated into their constituent atoms, which again scatter photons very efficiently (see section \ref{subsubsection:molecule creation}). This scheme has been employed to observe the loss of reactants in most studies of ultracold bi-alkali chemistry to date.

When it comes to studying the short-range chemistry, the quantum state specific nature of AMO techniques renders them ineffective, as the intermediate complex and the reaction products exist in a multitude of quantum states. Ion spectrometry, a tried-and-true method from the physical chemistry community, provides a universal and efficient way to detect all species involved in a reaction. Atoms or molecules involved in the reaction, regardless of their quantum states, can be ionized with UV photons of sufficiently high energy, and subsequently detected by time-of-flight (TOF) mass spectrometry. The velocity map imaging (VMI) technique \cite{eppink1997velocity} enables measurements of the kinetic energy distribution of the ions, from which valuable information on either the ionization process or the chemical reaction itself can be extracted. Additionally, ion detection has high sensitivity and low background noise, making it ideal for detecting intermediates and products which are much less concentrated compared to the reactants. It enabled the observation of a sample of ground state KRb molecules too dilute ($10^6 \textrm{cm}^{-3}$) to be optically imaged \cite{aikawa2010coherent}.

Thus, a comprehensive investigation of ultracold chemical reactions is best achieved by bringing together AMO and physical chemistry techniques. This is realized in the ``ultracold chemistry machine'' shown in Fig. \ref{figMachine}. The light blue shaded part of the apparatus can produce a mixture of ultracold, state-selected reactant atoms/molecules on-demand. As the reaction proceeds, the ion spectrometer in light orange shaded part of the apparatus can detect the reactants, products, and the transient intermediates with mass and quantum state sensitivity.

\subsection{Building the apparatus: divide and conquer}

Differences in AMO physics and physical chemistry techniques bring about many competing apparatus design requirements that needed to be resolved. Our strategy is to build interconnected vacuum chambers that each specializes in a set of mutually compatible tasks, and transfer the atoms and molecules in between them using electric, magnetic, and optical (laser) fields. One notable difference between an ultracold quantum gas apparatus and a gas-phase reaction dynamics apparatus (e.g. molecular beams) is the level of vacuum required. While the prior demands a pressure below $10^{-11}$ mbar due to the long experimental cycle time during which the collisions between the ultracold sample and background particles must be minimized, the latter only requires $ 10^{-6} $ to $ 10^{-8} $ mbar. The multi-chamber design allows the machine to be built and tested one vacuum chamber at a time. In the event that air-exposure is needed, the gate valves (G1, G2, and G3 in Fig. \ref{figMachine}) separating the chambers allows the negative impact on the vacuum to be localized.

\subsection{Production of ultracold KRb molecules} \label{subsection:KRb production}

\subsubsection{Atom cooling} \label{subsubsection:atom cooling}
 
The production of ultracold KRb begins with the loading and cooling of the precursor atoms. Bosonic $^{87}$Rb and fermionic $^{40}$K atoms are first loaded into a dual-species magneto-optical trap (MOT) in the ``MOT'' chamber, a Pyrex glass cell. At this stage, we usually have $1\times10^9$ Rb atoms (Fig. \ref{figMachine}(a)) and  $5\times10^6$ K (Fig. \ref{figMachine}(b)) atoms. We further increase the phase-space density (PSD) of the atoms by optically compressing the Rb MOT and applying gray molasses cooling to both species \cite{fernandes2012sub,rosi2018lambda}. We then optically pump the atoms to their respective stretched hyperfine states ($|F,m_F \rangle = |2,2\rangle$ for $^{87}$Rb and $|9/2,9/2\rangle$ for $^{40}$K) before capturing them in a quadrupole magnetic trap. This trap is formed by the ``transfer'' coil pair (Fig. \ref{figMachine}(c)), which operates in an anti-Helmholtz configuration and provides quadrupolar magnetic confinement for the atoms. The ``transfer'' coil, mounted on a meter-long transfer track, move the atoms through vacuum from the ``MOT'' chamber into the ``evaporation'' chamber. The ``MOT'' chamber is differentially-pumped with respect to the ``evaporation'' chamber through a thin tube (diameter = 10mm, length = 305mm) to maintain a relatively high vacuum ($<10^{-11}$mbar) inside the evaporation chamber. Once the atoms arrive at the evaporation chamber, they are loaded into a harmonic magnetic trap formed by three coils operated in a quadrupole-in-Ioffe configuration (QUIC) (Fig. \ref{figMachine}(e)) \cite{esslinger1998bose}. The QUIC trap has a minimum magnetic field value of 2.5G and trapping frequencies of $2\pi\times\{116, 116, 22\}$Hz for Rb and $2\pi\times\{171, 171, 33\}$Hz for K. Radio frequency (RF) evaporation~\cite{hess1986evaporative, petrich1995stable} is applied to the Rb atoms to reduce the temperature of the gas while the PSD is increased. K atoms remain in thermal equilibrium with Rb atoms throughout the evaporation process and therefore are sympathetically cooled. At the end of the RF evaporation we produce $4\times10^6$ Rb atoms and $1.5\times10^6$ K atoms at $\sim 2.5\mu\textrm{K}$. We then transfer the atoms from the QUIC trap into a focus-tunable optical dipole trap (TODT) formed by a single focused laser beam. The focal position of the TODT can be tuned smoothly over a distance of 30 cm by translating the position of a lens mounted on an air-bearing translation stage. Doing so transfers the atoms from the evaporation chamber to the center of the ``science'' chamber. Upon arrival, the atoms are loaded into a stationary ``crossed'' ODT (XODT) formed by crossing a ``horizontal'' beam (HODT) with a 30 $\mu$m beam waist and a ``vertical'' beam (VODT) with a 100 $\mu$m beam waist at near right angle. These two beams are frequency shifted from each other by 80 MHz to avoid interference. The arrangement of the ODT beams are illustrated in Fig. \ref{figMachine}. All of the ODT beams are generated by a single 1064 nm laser source with a linewidth of 1 kHz.

\subsubsection{Molecule creation} \label{subsubsection:molecule creation}

\begin{figure}[h]
\centering
\includegraphics[height=3.25 in]{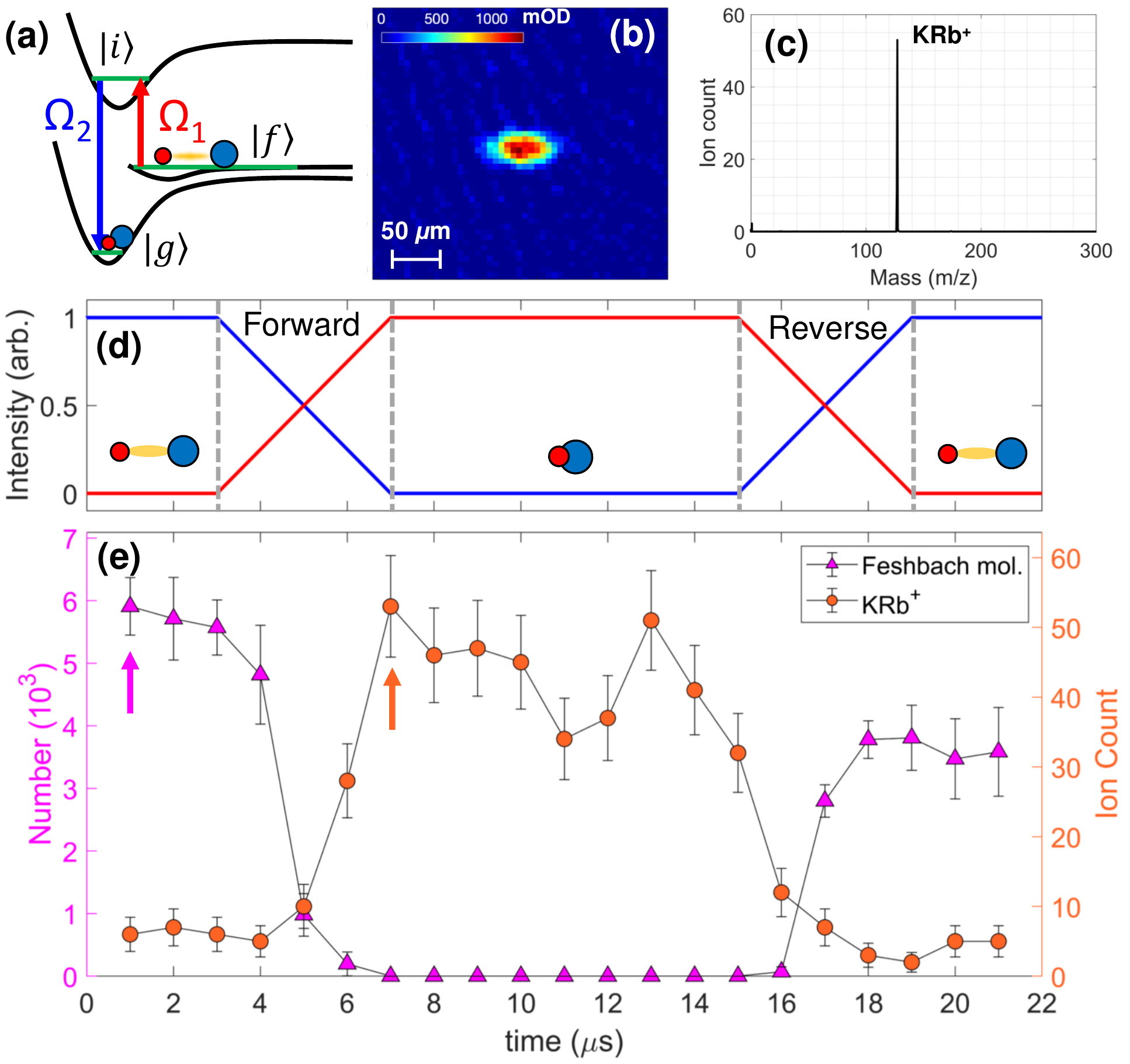}
    \caption{\textbf{Ground state molecule creation via STIRAP.} (a) Schematic diagram of STIRAP in KRb, showing how the 970 nm (red) and 690 nm (blue) lasers connect the $ |f\rangle, |i\rangle$ and $|g\rangle$ molecular states. the Rabi rates for the $ |f\rangle \rightarrow |i\rangle$ and $ |i\rangle \rightarrow |g\rangle$ transitions are $\Omega_1 = 4.3 \textrm{MHz}$ and $\Omega_1 = 5.5 \textrm{MHz}$, respectively. (b) An absorption image of $\sim 6000$ Feshbach molecules taken after the reverse STIRAP. The colorscale is in units of mOD (milli-optical-depth). (c) TOF mass spectrum showing the KRb$^+$ peak at $m/z = 127$. (d) The timing diagram of the (normalized) intensities of the 970 nm (red) and 690 nm (blue) lasers during a forward plus reverse STIRAP sequence. (e) The number of Feshbach molecules (magenta triangles) and KRb$^+$ ions (orange circles) at various times during the sequence in (d), showing the conversion of Feshbach molecules into ground state molecules and back. The data point indicated by the magenta (orange) arrow corresponds to the absorption image (mass spectrum) in (b)((c)).} 
\label{figSTIRAP}
\end{figure}

 To coherently associate the ultracold atoms into ultracold molecules, we take a two-step approach as demonstrated in Ref. \cite{ni2008high}. We first magneto-associate pairs of free Rb and K atoms into weakly-bond Feshbach molecules, then transfer the population into the rovibronic ground state via STImulated Raman Adiabatic Passage (STIRAP) \cite{bergmann1998coherent}. In preparation for the magneto-association, the Rb atoms are transferred from the $|2,2\rangle$ state into the $|1,1\rangle$ state via a microwave-driven adiabatic rapid passage (ARP), and the K atoms are transferred from the $|9/2,9/2\rangle$ state into the $|9/2,-9/2\rangle$ state via an RF-driven cascaded ARP. The ``FC'' coil pair around the ``science'' chamber is then switched on to produce a 550 G magnetic field and ODT evaporation is performed inside the XODT by lowering the intensity of the ``horizontal'' beam by a factor of $20$. This further increases the PSD of the atoms for the purpose of improving the efficiency of the magneto-association. At the end of the ODT evaporation we have $4\times10^4$ $^{87}$Rb atoms and $7\times10^4$ $^{40}$K atoms at $500\textrm{ nK}$ in the XODT. At this stage, the trapping frequencies are $2\pi\times\{265, 265, 60\}$ Hz for Rb atoms and $2\pi\times\{376, 376, 85\}$ Hz for K atoms. The magnetic field is then adiabatically ramped down from 550 G to 544 G across an inter-species Feshbach resonance centered at 546.62 G. This results in the creation of $\sim6\times10^3$ weakly-bound, Feshbach KRb molecules. To characterize the number and temperature of these Feshbach molecules, the magnetic field ramp is reversed to dissociate them back into free atoms, so that an absorption image of either atomic species can be taken. The image for Rb atoms is shown in Fig. \ref{figSTIRAP}(b).

The molecular population is then coherently transferred from the Feshbach state, $ |f \rangle $, into the $ |m_I^\text{K}, m_I^\text{Rb} \rangle = |-4, 1/2 \rangle$ hyperfine state of the lowest rovibronic state, $ |g \rangle = |X^1 \Sigma^+, v = 0, N = 0 \rangle$, using a pair of STIRAP laser pulses at 970 nm and 690 nm, with the electronically excited $ |i \rangle = |2^3\Sigma^+, v' = 23 \rangle$ state acting as an intermediate (see Fig. \ref{figSTIRAP} (a)). During the transfer, the intensities of the two pulses, which are proportional to the squared Rabi rates for the two transitions, are ramped according to the ``Forward'' part of Fig. \ref{figSTIRAP}(d). To characterize the ground state molecules using optical imaging, the STIRAP ramps are reversed and the back-converted Feshbach molecules are imaged. Fig. \ref{figSTIRAP}(e) shows the number of Feshbach molecules at various times during the STIRAP ramp sequence in Fig. \ref{figSTIRAP}(d). The integrated ion spectrometer (to be described in section \ref{subsection:ion spectrometer}) also allows us to directly probe the ground state KRb molecules using ionization followed by TOF mass spectrometry (see TOF mass spectrum in Fig. \ref{figSTIRAP}(c)). The evolution of the KRb$^+$ counts, which is proportional to the number of ground state molecules, shows a pattern complimentary to that obtained from imaging Feshbach molecules (see Fig. \ref{figSTIRAP}(e)). The reversible transfer of population between Feshbach and ground molecular states demonstrates the coherent nature of STIRAP. In a typical experiment, we obtain $\sim 5\times 10^3$ ground state KRb molecules at $500\textrm{ nK}$ with a peak density of $10^{12} \text{cm}^{-3}$. They are confined in the XODT with trapping frequencies of $2\pi\times\{300, 300, 68\}$ Hz. The molecular cloud is elliptically-shaped, with a characteristic size of $10 \times 10 \times 30 ~\textrm{um}$. The one-way STIRAP efficiency is $\sim 90\%$ for all experiments described in the rest of the paper \footnote{The one-way efficiency is defined as the square root of the round-trip efficiency, which is the percentage of Feshbach molecules recovered after the forward and reverse STIRAP sequence shown in Fig. \ref{figSTIRAP}(d). The STIRAP efficiency depends strongly on the relative coherence between the two laser frequencies. An improvement in this coherence resulted in an increase of the one-way efficiency from $\sim 70\%$, as shown by the measurement in Fig. \ref{figSTIRAP}(e), to the stated $\sim 90\%$.}.

For the purpose of studying the KRb + KRb reaction (Reaction \ref{KRb+KRb}), the remaining K and Rb atoms that did not form molecules are pushed out of the XODT using 5 ms long resonant light pulses. The amount of atoms that survive the pushout is measured with ion TOF mass spectrometry to be less than 200 for each species. For future studies of the K/Rb + KRb reaction (Reactions \ref{K+KRb} and \ref{Rb+KRb}), we can leave a variable amount of atoms in the ODT by adjusting the intensity and duration of the atom push-out pulse.

The typical cycle time from the loading of the dual species MOT to the production of ground state molecules is 50 s, with MOT loading (6 s), RF evaporation (30 s), and ODT evaporation (5 s) accounting for the majority of the cycle time. This relatively long time scale is typical for ultracold quantum gas experiments, which share many of the same steps \footnote{Much faster cycle times can be achieved for experiments with single alkali species~\cite{rudolph2015high,urvoy2019direct}\. The techniques involved, however, cannot be easily applied to a bi-alkali mixture experiment such as ours.}.

\subsection{Probing ultracold reactions: the ion mass and kinetic energy spectrometer} \label{subsection:ion spectrometer}

As soon as the ultracold reactants are created in the XODT, chemical reactions proceed continuously through two-body collisions. This is in contrast to molecular beam experiments, the traditional platform for studying chemical dynamics, where in each iteration of the experiment all the reactions occur over a narrow time window, defined by either an initializing laser pulse or the crossing between two molecular beams. As the reaction progresses in the trap, we probe it by ion mass and kinetic energy spectrometry. Neutral species from the reaction are first photoionized by a UV laser pulse, then accelerated by VMI ion optics, and finally detected by a multi-channel plate (MCP) detector. The MCP records the TOFs and hit locations of the ions, from which mass and kinetic energy information can be respectively extracted. In the subsections below, we discuss the design of key components of the spectrometer.

\subsubsection{Ionization source} \label{subsubsection:ionization source}

The ultracold reactions we probe involve a variety of chemical species with distinct photoionization characteristics (Fig. \ref{figEnergyLevel}). To efficiently ionize each species, a frequency tunable source is desired. TOF mass spectrometry requires a well-defined time-zero, which demands a pulsed source. To fulfill the above two requirements, we use a frequency-doubled, broadly tunable dye laser (LIOP-TEC/LiopStar-HQ) pumped by a pulsed Nd:YAG laser (EdgeWave BX80) as our ionization source. The system has a pulse duration of 7 ns, a spectral width of 0.06 cm$^{-1}$, and a tuning range of 220 - 400 nm (after frequency-doubling by a BBO3 crystal). The wavelength of the laser is calibrated to within 0.02 cm$^{-1}$ using a laser wavelength meter.

The requirement for a pulsed source is somewhat at odds with the continuous nature of our reactions. As the reaction progresses, both the intermediate complex and the products are continuously produced and then quickly ``lost'' by either dissociation or escaping from the detection region. Since the cycle time of our experiment is long and the particle number low, it is important to maximize the total number of products/intermediates that can be ionized in each experimental cycle in order to achieve a reasonable data collection rate. This number scales linearly as both the pulse energy and the repetition rate, i.e. the time-averaged power, of the ionization source. Very high pulse energies can result in undesirable effects such as the saturation of ionization probability, space-charge effects, and two-photon processes. Therefore, an ionization source with a high repetition rate and a moderate pulse energy is desired. An upper limit on the repetition rate is imposed by the range of TOFs expected for the ions of interest, which in our case can be as long as 130 $\mu$s. we therefore chose an ionization laser with a tunable repetition rate of up to 10 kHz and a pulse energy of 10 - 100 $\mu$J.

\subsubsection{Photoionization scheme} \label{subsubsection:ionization strategy}

\begin{figure}[h]
\centering
\includegraphics[height=5 in]{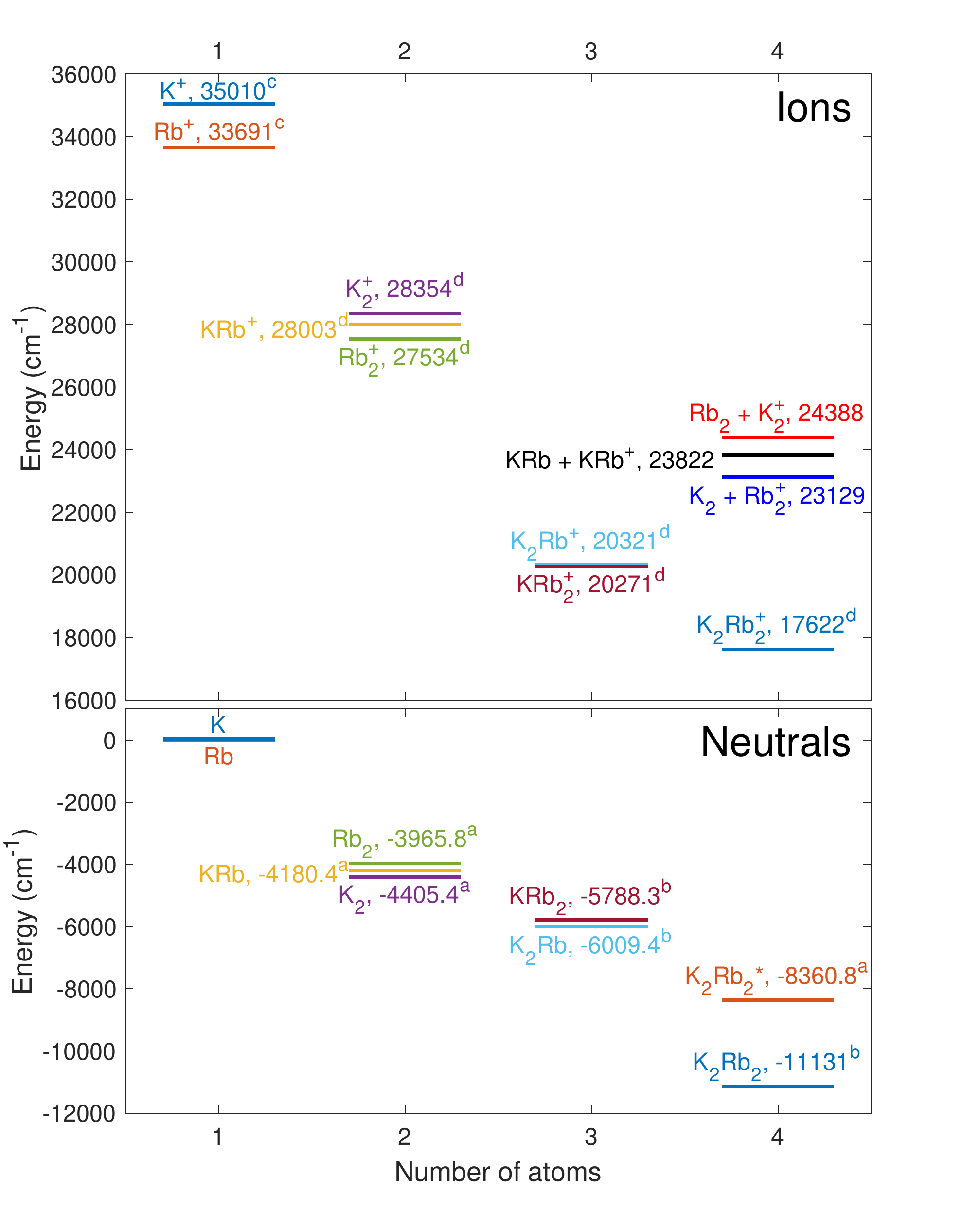}
    \caption{Ground state energies of neutral (lower panel) and ionic (upper panel) species relevant to the ultracold reactions involving KRb molecules (reactions \ref{KRb+KRb}, \ref{K+KRb}, and \ref{Rb+KRb}). The species are sorted horizontally by the number of constituent atoms. The energy of the K$_2$Rb$_2^*$ complex is equal to that of two separate KRb molecules. \\
    $^a$ spectroscopic data, Ref. \cite{ospelkaus2010quantum}. \\
    $^b$ \textit{ab initio} calculation, Ref. \cite{byrd2010structure}.\\
    $^c$ spectroscopic data, Ref. \cite{sansonetti2006wavelengths}\cite{sansonetti2008wavelengths}.\\
    $^d$ \textit{ab initio} calculation, Ref. \cite{hu2019direct}.\\}
\label{figEnergyLevel}
\end{figure}

\begin{figure*}
\centering
\includegraphics[width=6.5 in]{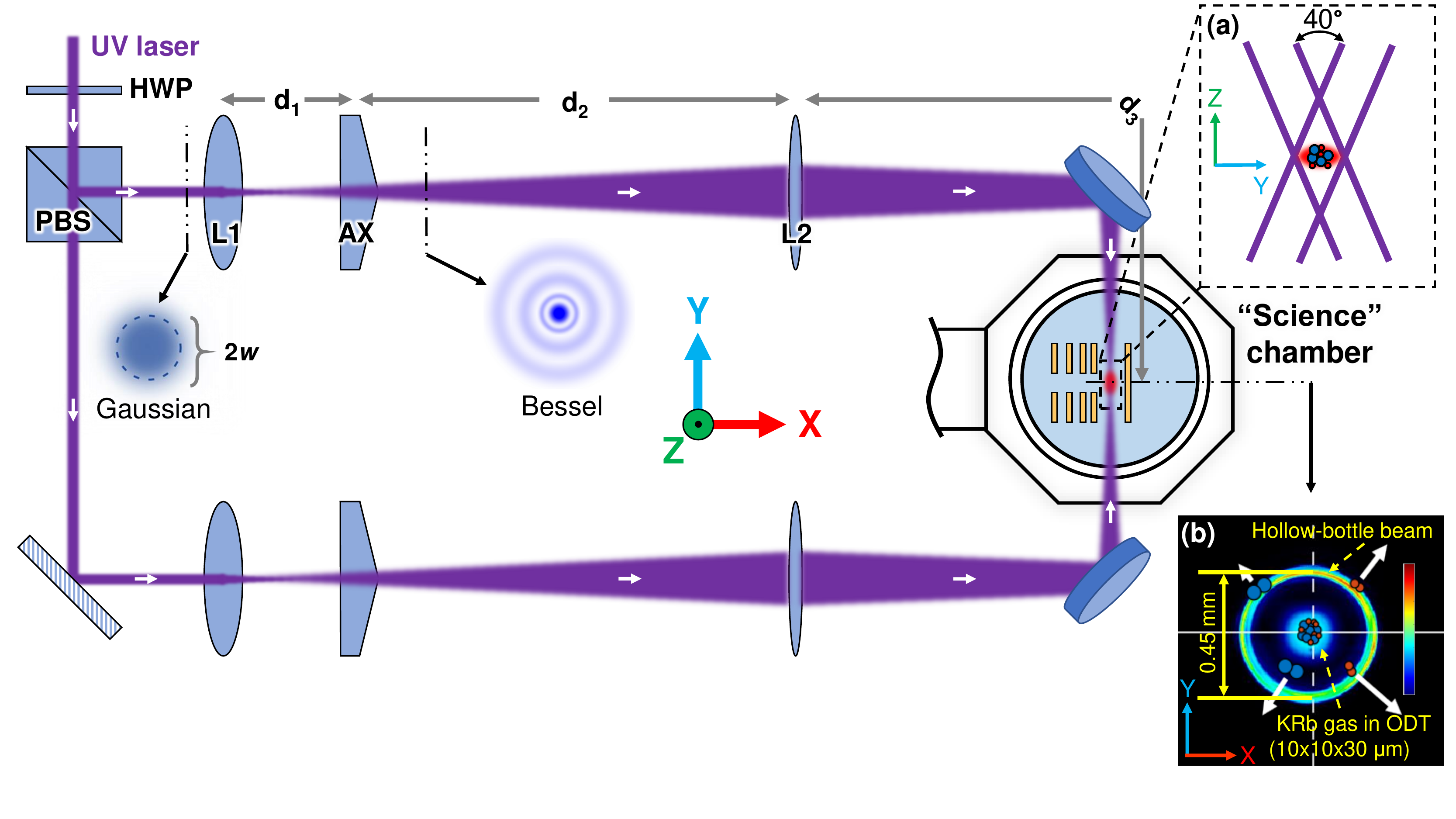}
    \caption{Crossed hollow-bottle beam optical setup. HWP: half-wave plate; PBS: polarizing beam-splitter; L1: bi-convex lens (effective focal length 283 mm). AX: axicon (physical angle 1$^\circ$); L2: bi-convex lens (effective focal length 46.2 mm); $d_1 = 80 $ mm ; $d_2 = 670 $ mm; $d_3 = 472$ mm. (a) Schematic cross-sectional view in the $yz$ plane showing the relative placement of the pair of crossed hollow beams with respect to the reactant cloud. (b) Schematic cross-sectional view in the $xy$ plane with the intensity profiles of the hollow-bottle beams and the VODT beam superimposed; the reactant cloud is located at the center of the VODT. A fraction of the products flying away from the reactants are intercepted and ionized by the hollow-bottle beams.} 
\label{figHollowBeam}
\end{figure*}

Below we outline the strategy for ionizing the reactants, the intermediates , and the products, respectively. The ground state energies of all relevant species and their first-ionized counterparts are given in Fig. \ref{figEnergyLevel}.

All \textit{reactants} (K, Rb, and KRb) can be ionized by 285 nm or shorter wavelengths. We shape the ionization laser into a Gaussian beam and center it around the reactant cloud. Due to the high density of reactants, the UV pulse energy must be kept low to ensure that no more than one ion is generated per pulse to prevent ion count saturation and space-charge effects.

When ionizing \textit{products} (K, K$_2$, and Rb$_2$), we must consider the fact that some of them have higher $IP$ compared to the reactants, and that the same photons that ionize them can ionize and deplete the reactants, leading to a reduction in the total number of products generated and detected per cycle. To circumvent this issue, we shape the ionization beam into a hollow-bottle beam with a ring profile around the reactant cloud (Fig. \ref{figHollowBeam}(b)). The products will have enough kinetic energy to escape the shallow trapping potential ($\sim 10 \mu$K ) of the ODT and reach the ionization beam. Reactants, on the other hand, are left mostly in the dark.

The optical setup we use to create the hollow-bottle beam \cite{Kulin_2001} is shown in Fig. \ref{figHollowBeam}. The combination of two lenses and an axicon (L1, L2, and AX in Fig.\ref{figHollowBeam}) transforms the input Gaussian beam first into a Bessel beam, and finally into a hollow-bottle beam. The focal plane of the bottle beam, where the ring becomes the sharpest (0.45 mm diameter, 5.4 $\mu$m Gaussian ring width), intercepts the reactant cloud (Fig. \ref{figHollowBeam}(b)). The bottle closes up at 6.4 mm and 16.5 mm away from either side of its focus. We generate two such beams and cross them at a 40$^\circ$ angle around the reactant cloud (Fig. \ref{figHollowBeam}(a)). We measure a 500:1 contrast ratio between the intensities at (the peak of) the ring and the center of the beam. At the UV pulse energy and repetition rate we typically use to ionize the products, we observe an acceptable amount of depletion of the reactants (see section \ref{section: product and intermediate results}).

\textit{Intermediates} (K$_2$Rb$^*$, KRb$_2^*$, and K$_2$Rb$_2^*$) are formed by the collision between a pair of reactants and are therefore present inside the ODT. To ionize them we use a Gaussian beam profile similar to that used to ionize reactants. Fortunately, all the intermediates have lower IP than the lowest of the reactants (KRb, $IP =  3.23 \times 10^4 ~\textrm{cm}^{-1}$), and therefore the ionization of the intermediates does not result in significant depletion of the reactants (see section \ref{section: product and intermediate results}).

\subsubsection{VMI ion optics} \label{subsubsection:VMI ion optics}

\begin{figure*}
\centering
\includegraphics[width=6.5in]{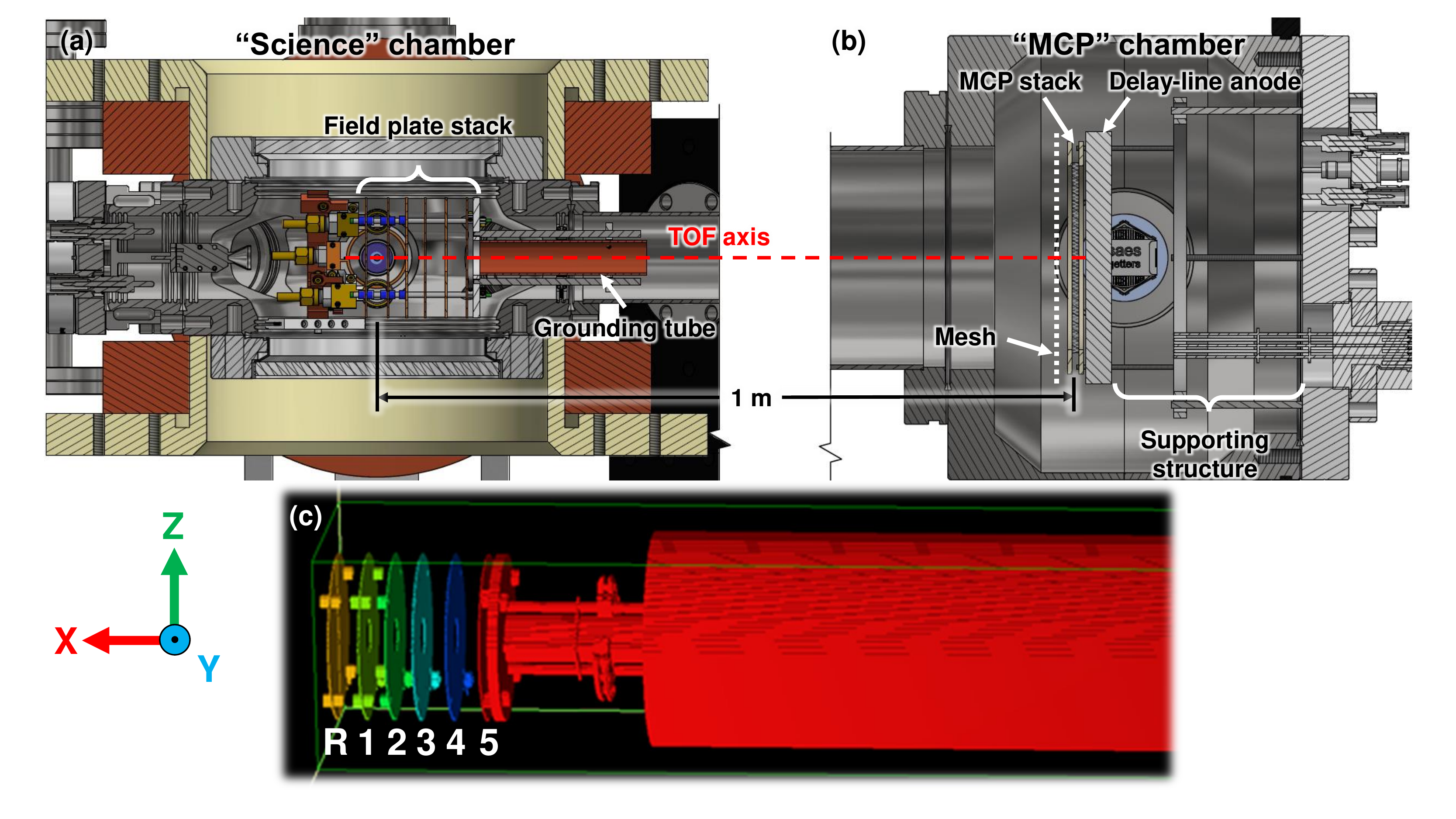}
\caption{\textbf{The ion spectrometer.}  (a)(b) CAD drawing showing VMI ion optics and MCP detector situated inside the ``science'' and ``MCP'' chambers, respectively. (c) A realistic 3D model of the ion optics is imported into SIMION for geometry optimization. The six electric field plates are labeled R, 1, 2, 3, 4, and 5. The final design parameters are summarized in Tab.~\ref{tab:VMI}.}
\label{figVMI}
\end{figure*}

Kinetic energy spectrometry can be achieved through ion imaging, a workhorse technique behind many studies of gas-phase chemical dynamics~\cite{ashfold2006imaging}. In an experiment, ions of a given species and kinetic energy $ KE $ are distributed on a Newton sphere. This sphere expands as the ion packet, accelerated by an electric field, flies towards the detector along the TOF axis. By the time the sphere reaches the detector, it will have expanded to a radius $ R $ that is related to $ KE $ according to

\begin{equation}
    R = A\cdot\sqrt{KE/V_R},
    \label{eqnVMI}
\end{equation}

where $ A $ is a proportionality constant that depends on the distance of flight, the mass of the ion, and details of the acceleration E-field. The Newton is projected onto the detector to form a 2D image, from which its radius $R$ can be extracted through either mathematical reconstruction or time-slicing methods \cite{ashfold2006imaging}. Time-slicing is enabled in our experiment by the delay-line MCP's ability to record both the time-of-flight and the position of each ion hit. This allows a narrow time slice in the middle of the Newton sphere, typically $\sim 10\%$ of the total ion packet duration, to be selected and analyzed for $R$.

Electric field plates are typically employed to apply the acceleration E-field to the ions. While a simple two-plate design could provide mapping between an ion's velocity and position on the detector, it also inevitably maps its initial location in the ionization volume onto the position on the detector. The kinetic energy resolution of such a design is ultimately limited by the initial spatial extent of the ion distribution. This issue can be circumvented by employing VMI ion optics, which consists of multiple field plates with carefully designed geometries and voltages. Such a configuration creates an E-field in which a one-to-one mapping between the velocity of the ions and their position on the detector is achieved.

\begin{table}[b]
\small
\caption{\label{tab:VMI} Dimensions of the electric field plates consisting the VMI ion optics. All plates have an outer diameter of 35 mm and are 0.81 mm thick. }
  \begin{tabular*}{0.48\textwidth}{lp{0.10\textwidth}lp{0.10\textwidth}lp{0.10\textwidth}lp{0.12\textwidth}}
\hline
Electrodes & Aperture diameter (mm) & Relative voltage (\%) & Distance to previous electrode (mm)\\
\hline
R & 3.0 & 100.0 & \\
1 &11.0 & 82.0 & 9.19\\
2 &12.5 & 58.5 & 7.70\\
3 & 14.0 & 35.0 & 7.70\\
4 &12.5 & 17.5 & 8.71\\
5 &11.0 & 0.0 & 8.71\\ \hline
\end{tabular*}
\end{table}

\begin{figure*}[ht!]
\centering
\includegraphics[width = 7.25 in]{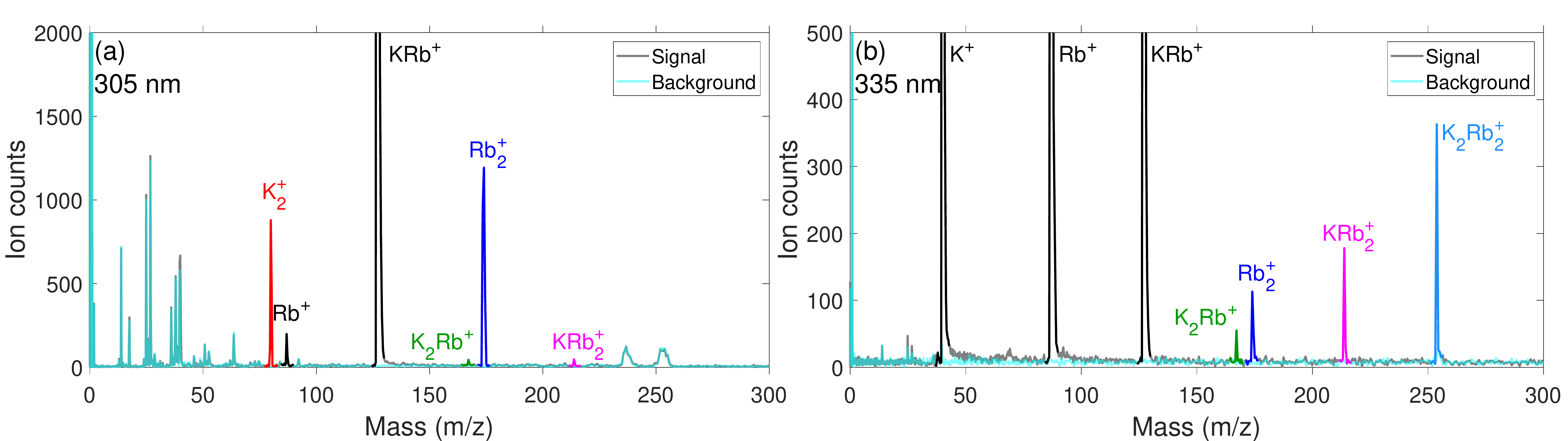}
\caption{TOF mass spectra acquired at an ionization laser wavelength of (a) 305 nm and (b) 335 nm, respectively. The 305 nm spectrum is adapted from Figure S4 of Ref. \cite{hu2019direct}. Gray and cyan traces correspond to ``signal'' and ``background'' spectra, recorded in the presence of absence of atoms and molecules in the ODT, respectively. Mass peaks that correspond to combinations of K and Rb atoms are highlighted in color. Spectrum (a) shows the detection of K$_2^+$ and Rb$_2^+$ ions, resulting from the photoionization of the reaction products K$_2$ and Rb$_2$. Spectrum (b) shows the detection of K$_2$Rb$^+$, KRb$_2^+$ ions resulting from the dissociative ionization of the transient intermediate K$_2$Rb$_2^*$, and K$_2$Rb$_2^+$ ions from the direct ionization of  K$_2$Rb$_2^*$. The origins of the other mass peaks in the spectra are discussed in the main text (section \ref{section: product and intermediate results}).}
\label{figMassSpec}
\end{figure*}

Our VMI setup design is guided by Ref. \cite{eppink1997velocity} and \cite{leon2014design}. The CAD drawing of the VMI ion optics is shown in Fig \ref{figVMI}(a). The electric field plate stack consists of five gold-plated copper plates (labeled from left to right as R, 1,2,3,4,and 5) supported by four alumina rods and spaced from each other by alumina spacers. The reactant cloud is placed at the center between the repeller (R) and extractor (1) plates. To determine the optimal geometries and voltages of the five plates, we use SIMION to simulate VMI performance. The optimization parameters include the aperture size of each plate, the distances between neighboring plates, and the voltages of the electrodes. We imported our 3D mechanical model into SIMION (v8.1) to perform an accurate simulation (Fig. \ref{figVMI}(c)). The final design parameters are summarized in Tab. \ref{tab:VMI}. The voltage applied to each plate is reported as a percentage of the repeller voltage $V_R$. An appropriate $V_R$ must be chosen for each chemical reaction we study to achieve a balance between kinetic energy resolution and detection range, given the finite size of the MCP and the apertures along the TOF axis. Two primary settings are used in our experiments. In the ``low energy'' (``high energy'') setting, $V_R$ is set to be around 100 V (1000 V), which gives an energy range of $\sim 0 - 150 \textrm{cm}^{-1}$ ($\sim 0 - 2500 \textrm{cm}^{-1}$) and is appropriate for studying the product kinetic energy distribution of the KRb + KRb (K + KRb) reaction for both ground and low-lying rovibrationally excited reactants. The proportionality constant $ A $ in Eq. \ref{eqnVMI} is $16.44 \textrm{mm} / \sqrt{\textrm{cm}^{-1}/\textrm{V}}$ from simulation. Experimental calibration of $ A $ is discussed in section \ref{section: VMI calibration}. Imperfections in the real experimental setup limits the energy resolution of VMI (the ability to distinguish adjacent VMI rings). For ``low energy'' setting, we find the minimum resolvable energy difference to be 0.1 cm$^{-1}$ around 0 cm$^{-1}$, 2 cm$^{-1}$ around 10 cm$^{-1}$, and 6 cm$^{-1}$ around 100 cm$^{-1}$. The resolution for the ``high energy'' setting is overall $\sim 3.3$ times lower.

\subsubsection{Detector} \label{subsubsection: MCP detector}

For detecting the ions we use a delay-line MCP (Roentdek DLD80) (Fig. \ref{figMachine}(e), Fig. \ref{figVMI}(b)). The MCP has an active diameter of 80 mm, a spatial resolution of 0.08 mm, and a temporal resolution of 1 ns. Details of the detector performance can be found in Ref. \cite{jagutzki2002broad}. The VMI voltages that we typically use (100 - 1000 V) is insufficient to accelerate the ions to high enough kinetic energies needed for high MCP detection efficiency. To improve the efficiency, we place a grounded stainless-steel mesh 3 mm before the MCP stack front surface and bias it to -3.8 kV to create an extra stage of acceleration. The resulting detection efficiency, after accounting for the mesh transmission (75$\%$), the MCP open-area-ratio (60$\%$), and the mass-dependent intrinsic MCP efficiency (calculated based on an empirical formula developed in \cite{krems2005channel}), is 0.394 (0.282) for the lightest (heaviest) ion, K$^+$ (K$_2$Rb$_2^+$), in our experiment.

\begin{figure*}[ht!]
\centering
\includegraphics[width=7.25 in]{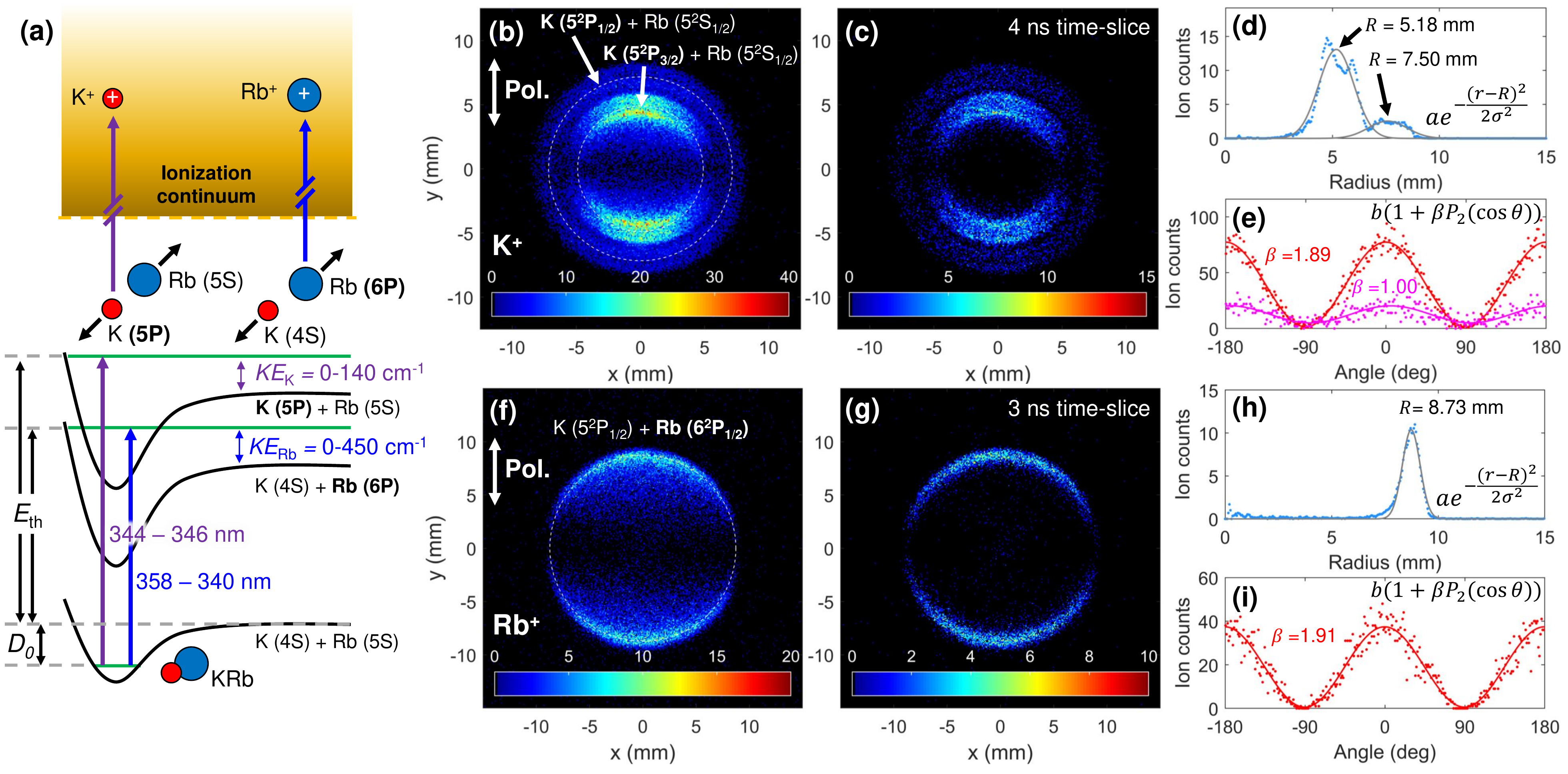}
\caption{One-color, two-photon dissociation plus ionization of ultracold KRb (a) Schematic. The first photon dissociates KRb into photofragments with well-defined kinetic energies, and the second photon ionizes the excited fragment (marked in bold). The values for $D_0$ and $E_{\textrm{th}}$ can be found in Tab. \ref{tab:KRbPD}. (b) Ion image (projected) obtained by monitoring K$^+$ from KRb PD plus ionization with $ V_R = 100.2 $ V, $ \lambda_{\textrm{UV}} = 345.8103 $ nm, and vertical laser polarization. The inner and outer rings correspond to the K (5$^2$P$_{3/2}$) + Rb (5$^2$S$_{1/2}$) and K (5$^2$P$_{1/2}$) + Rb (5$^2$S$_{1/2}$) PD channels, respectively. The rings are broadened due to photoionization recoil (0.253 cm$^{-1}$). (c) Image of a 4 ns time-slice of the entire 40 ns ion packet. (d) The radial intensity distribution of the time-slice. Gaussian fits are applied to extract the radii ($R$) of the two Newton spheres, which are $5.18 \pm 0.03 $ mm and $7.50 \pm 0.05 $ mm, respectively. (e) The angular intensity distributions of the inner (red) and outer (magenta) rings in the time-slice. The distributions are fitted to the function $A(1 + \beta P_2(\cos \theta))$. The the resulting anisotropy parameter $\beta$ is $1.89 \pm 0.05 $ for the inner ring, indicating the PD transition is primarily ``parallel'', and $1.00 \pm 0.08 $ for the outer ring, indicating a mix of ``parallel'' and ``perpendicular'' characters. (f) Ion image (projected) obtained by monitoring Rb$^+$ from KRb PD plus ionization with $ V_R = 992 $ V, $ \lambda_{\textrm{UV}} = 346.0789 $ nm, and vertical laser polarization. The ring corresponds to the K (4$^2$S$_{1/2}$) + Rb (6$^2$P$_{1/2}$) PD channel, and is broadened due to photoionization recoil (0.125 cm$^{-1}$). (g) Image of a 3 ns time-slice of the entire 30 ns ion packet. (h) The radial and (i) angular distributions of the time-slice. $R = 8.73 \pm 0.01$ mm and $\beta = 1.91 \pm 0.07$.}
\label{figPDAnalysis}
\end{figure*}

\section{Detecting the products and the intermediate complex of the KRb + KRb reaction } \label{section: product and intermediate results}

We used ion mass spectrometry to directly probe for the products and the intermediate complex of the ultracold KRb + KRb reaction (reaction \ref{KRb+KRb}). For these experiments, $\sim 5000$ KRb molecules were prepared in the rovibronic ground state. As the sample reacts, the products (intermediate complex) were ionized using a crossed hollow-bottle (Gaussian) UV beam profile with 305 (335) nm wavelength, 60 (14) $\mu$J average pulse energy, and 3 (7) kHz repitition rate. Ion signals were accumulated for 800 (1200) experimental cycles to result in the TOF mass spectra shown in Fig. \ref{figMassSpec}(a) (Fig. \ref{figMassSpec}(b)). Each ``signal'' spectrum (gray) is accompanied by a ``background'' spectrum (cyan), recorded without atoms and molecules in ODT, to distinguish mass peaks that correspond to species of interest from noise peaks. The detection of the products and intermediate complex were first presented in Ref. \cite{hu2019direct}, where ionization was carried out using different wavelengths. In the 305 (335) nm case, single-photon (two-photon) ionization of the reactant KRb molecules results in a $30\%$ ($20\%$) depletion of their population over the course of the reaction, and a strong KRb$^+$ ($m/z = 127$) ion signal in the spectrum.

From the 305 nm mass spectrum, we identify mass peaks that correspond to K$_2^+$ ($m/z = 80$) and Rb$_2^+$ ($m/z = 174$), resulting from single-photon ionization of the reaction products K$_2$ and Rb$_2$. The residual light intensity at the center of the crossed hollow beams results in a small but observable amount of dissociative ionization of species present inside the XODT. In particular, the pathway $ \text{KRb} \xrightarrow{h\nu} \text{KRb}^+ \xrightarrow{h\nu} \text{K} + \text{Rb}^+ $ leads to the Rb$^+$ ($m/z = 87$) signal, and the pathways $ \text{K}_2\text{Rb}_2^* \xrightarrow{h\nu} \text{K}_2\text{Rb}^+ + \text{Rb} $ and $ \text{K}_2\text{Rb}_2^* \xrightarrow{h\nu} \text{K}\text{Rb}_2^+ + \text{K} $ lead to the K$_2$Rb$^+$ ($m/z = 167$) and KRb$_2^+$ ($m/z = 214$) signals, respectively.

The 335 nm mass spectrum shows mass peaks that correspond to K$_2$Rb$^+$ ($m/z = 254$), KRb$_2^+$, and K$_2$Rb$_2^+$, resulting from the pathways $ \text{K}_2\text{Rb}_2^* \xrightarrow{h\nu} \text{K}_2\text{Rb}^+ + \text{Rb} $, $ \text{K}_2\text{Rb}_2^* \xrightarrow{h\nu} \text{K}\text{Rb}_2^+ + \text{K} $, and $ \text{K}_2\text{Rb}_2^* \xrightarrow{h\nu} \text{K}_2\text{Rb}_2^+ $, respectively. As demonstrated in our previous result \cite{hu2019direct}, once the ionization photon energy drops below the lowest two dissociation thresholds of K$_2$Rb$_2^+$ which lead to triatomic ions (e.g. 356 nm), the single-photon pathway is the only one that remains. Two-photon ionizations of KRb ($ \text{KRb} \xrightarrow{2h\nu} \text{KRb}^+ $) and Rb$_2$ ($ \text{Rb}_2 \xrightarrow{2h\nu} \text{Rb}_2^+ $) resulted in the KRb$^+$ and Rb$_2^+$ ion signals in the spectrum. The weak nature of these processes is compensated by the high density of KRb and Rb$_2$ present inside the XODT to yield observable signals. Two-photon dissociation plus ionization of KRb results in the K$^+$ and Rb$^+$ signals. We take advantage of this process for the calibration described in section \ref{section: VMI calibration}.

\section{Calibration of the kinetic energy spectrometer} \label{section: VMI calibration}

Calibrating the kinetic energy spectrometer means experimentally determining the relationship between the radii of ion Newton spheres and their associated kinetic energies. This requires the generation of ions with sufficient and well-known kinetic energies, which is a challenge for ultracold systems where particles move with negligible kinetic energies. Photodissociation (PD) of diatomic molecules, which imparts a well-defined amount of kinetic energy into the recoiling photofragments, provides a suitable solution to this challenge.\cite{eppink1997velocity} We used the one-color, two-photon dissociation plus ionization process shown schematically in Fig. \ref{figPDAnalysis} (a). The first photon dissociates the ground state KRb molecules into K and Rb photofragments, and the second photon ionizes the optically excited fragment. Ion images were acquired at different photon energies. For each ion image, $\sim$ 5000 KRb molecules in their rovibrational ground state were exposed to 5000 UV laser pulses over 1 s for 500 - 1500 experimental cycles. The use of ultracold precursors in a single quantum level for PD ensures the production of photofragments with sharply-defined kinetic energies, a feature especially beneficial to calibration at low kinetic energies. The average number of ions generated per UV pulse is kept much less than one to ensure the resulting ion images not broadened by space-charge. Separate calibrations are carried out for the ``low energy'' and ``high energy'' repeller voltage ($ V_R $) settings.

\begin{table}[t]
\small
\caption{\label{tab:KRbPD} Summary of relevant energies for KRb photodissociation and ionization. All values are relative to the $^{40}$K $^2$4S$_{1/2}$ + $^{87}$Rb $^2$5S$_{1/2}$ asymptote at zero energy. $D_0$: dissociation energy; $E_{\textrm{th}}$: atomic threshold energy; $IP$: ionization potential.}
\begin{tabular*}{0.48\textwidth}{@{\extracolsep{\fill}}lll}
\hline
Quantity & Energy (cm$^{-1})$ & Reference \\
\hline
$ D_0$ ($^{40}$K$^{87}$Rb) & -4180.42 & \cite{ospelkaus2010quantum} \\
$ E_{\textrm{th}} $ ($^{40}$K 5$^2$P$_{1/2}$ + $^{87}$Rb 5$^2$S$_{1/2}$) & 24701.38 &  \cite{mckay2011low}\\
$ E_{\textrm{th}} $ ($^{40}$K 5$^2$P$_{3/2}$ + $^{87}$Rb 5$^2$S$_{1/2}$) & 24720.13 &  \cite{mckay2011low}\\
$ E_{\textrm{th}} $ ($^{40}$K 4$^2$S$_{1/2}$ + $^{87}$Rb 6$^2$P$_{1/2}$) & 23715.08 &  \cite{sansonetti2006wavelengths}\\
$ E_{\textrm{th}} $ ($^{40}$K 4$^2$S$_{1/2}$ + $^{87}$Rb 6$^2$P$_{3/2}$) & 23792.59 &  \cite{sansonetti2006wavelengths}\\
$ IP(^{40}\textrm{K}) $ & 35009.81 & \cite{sansonetti2008wavelengths} \\
$ IP(^{87}\textrm{Rb}) $ & 33690.81 & \cite{sansonetti2006wavelengths} \\
\hline
\end{tabular*}
\end{table}

For ``low energy'' calibration, $V_R$ is set to 100.2 V. We photodissociated KRb into K (5P) and Rb (5S) and ionized K (5P) to produce $\textrm{K}^+$ ions with kinetic energy ($KE$) in the 0 - 140 cm$^{-1}$ range (Fig. \ref{figPDAnalysis} (a)) by varying the UV laser wavelength (photon energy) between 346 nm (29802 cm$^{-1}$) and 344 nm (29070 cm$^{-1}$). A 5G magnetic quantization field is maintained for the molecules \footnote{As demonstrated in Ref. \cite{ospelkaus2010quantum}, maintaining the quantization to the fermionic KRb molecules is essential to keeping them in a single quantum state, such that they undergo $p$-wave collision and the sample enjoys a relatively long lifetime. A longer-lived sample allows us to collect more ions towards the calibration per experimental cycle.}. An example K$^+$ ion image acquired at a UV wavelength of 345.7325 nm is shown in Fig. \ref{figPDAnalysis} (b). From the image we identify two anisotropic ring patterns, and assign the inner and outer rings to ions dissociated from the K (5$^2$P$_{3/2}$) + Rb (5$^2$S$_{1/2}$) and the K (5$^2$P$_{1/2}$) + Rb (5$^2$S$_{1/2}$) channels, respectively. We note that the rings are broadened due to a 0.253 cm$^{-1}$ of recoil energy imparted to the K$^+$ ions by the ionizing UV photon. To determine the radii of the two underlying Newton spheres, we select ions from the central 4 ns time-slice of a 40 ns distribution along the TOF axis (Fig. \ref{figPDAnalysis} (c)). The time-slice is then angle-integrated to obtain the radial intensity distribution (Fig. \ref{figPDAnalysis} (d)). Gaussian fits are applied to the radial distribution to extract the radii $R$, which are $5.18 \pm 0.03$ mm and $7.50 \pm 0.05$ mm for the inner and outer Newton spheres respectively. The anistoropy of each Newton sphere is analyzed by examining the angular intensity distribution (Fig. \ref{figPDAnalysis} (e)) of the corresponding ring in the time-slice. The distributions are fitted to the function $b(1 + \beta P_2(\cos \theta))$ \footnote{This formula results from a quasiclassical model describing the angular distribution of fragments produced by the photodissociation of molecules prepared in spherically symmetric states\cite{zare1963doppler}, \textit{e.g.} KRb prepared in the rovibrational ground state. Here $b$ is a constant, $\beta$ is the anisotropy parameter that varies from +2 (``parallel'' transition) to -1 (``perpendicular'' transition), $\theta$ is the angle in the plane of the image, and $P_2$ is the second order Legendre polynomial.}. The resulting anisotropy parameter $\beta$ is $1.89 \pm 0.05 $ ($\approx 2$) for the inner ring, indicating that the PD transition dipole moment is mostly parallel to the molecular axis. For the outer ring, $\beta = 1.00 \pm 0.08 $, indicating a mix of parallel of perpendicular characters in the PD transition. Detailed understanding of the observed photofragment anisotropy in Fig. \ref{figPDAnalysis} is a subject for future study.

\begin{figure} [t!]
\centering
\includegraphics[height=3.85 in]{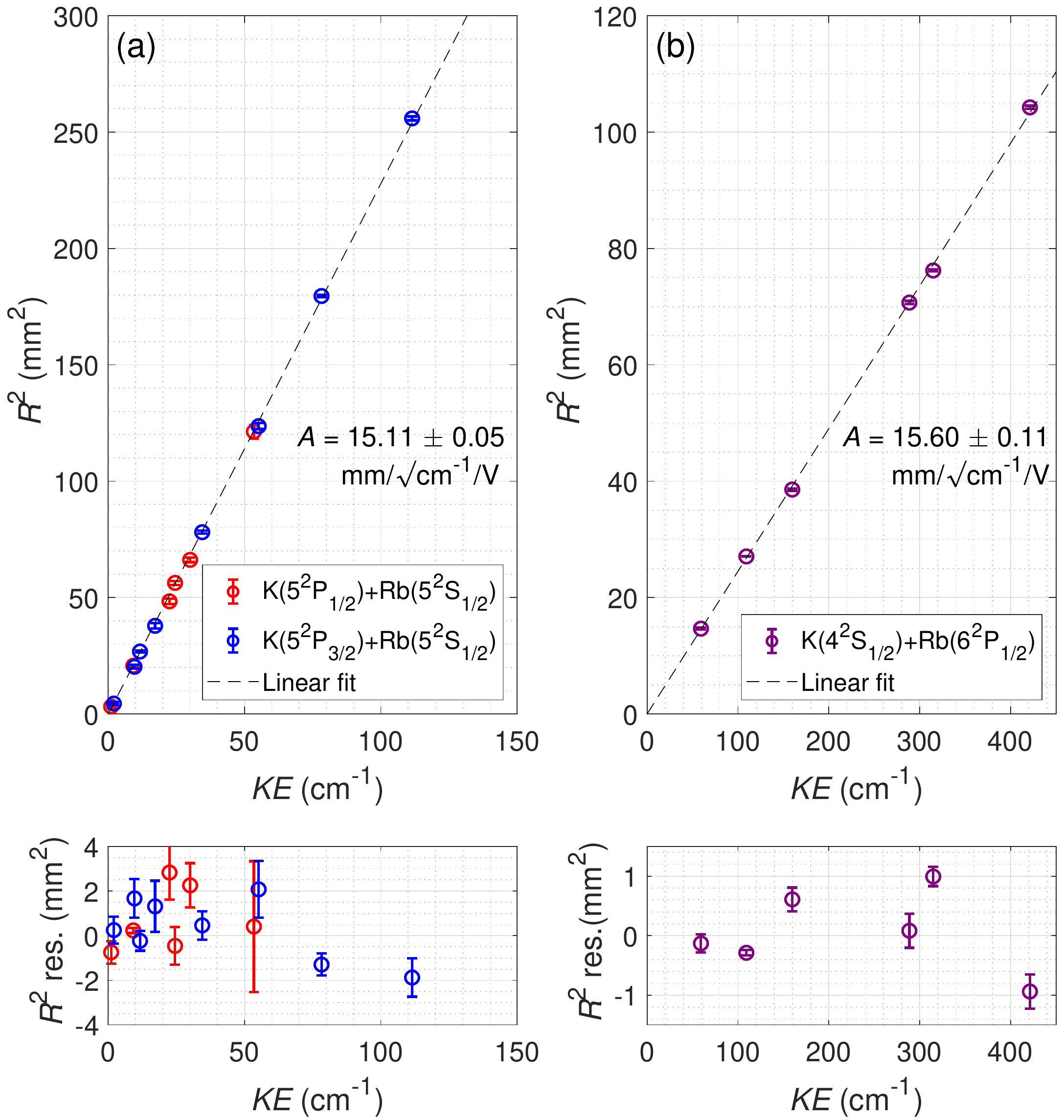}
\caption{\textbf{Calibrating the kinetic energy spectrometer} (a) Top panel: calibration data and fit to $ R^2 = A^2 KE /V_R $ for the ``low energy'' setting. Contributions from the K (5$^2$P$_{3/2}$) + Rb (5$^2$S$_{1/2}$) and K (5$^2$P$_{1/2}$) + Rb (5$^2$S$_{1/2}$) channels are colored blue and red, respectively. Bottom panel: residual of the fit. (b) calibration data and fit for the ``high energy'' setting. Bottom panel: residual of the fit. The values of the calibration constant $A$ obtained from the fits are close to the designed value of $16.44 \textrm{mm} / \sqrt{\textrm{cm}^{-1}/\textrm{V}}$.}
\label{figVMICal}
\end{figure}

The kinetic energy of ions associated with each Newton sphere is given by

$$ KE = \frac{m_{ \textrm{Rb}}}{m_{\textrm{Rb}} + m_{\textrm{K}}} \left( E_{\gamma} - E_{\textrm{th}} + D_0 \right),$$

where $ E_{\gamma} $ is the photon energy of the laser (calibrated to within 0.02 cm$^{-1}$), $ E_{\textrm{th}} $ is the atomic threshold energy of the corresponding PD channel, and $ D_0 $ is the dissociation energy of the rovibronic ground state KRb molecule (see Tab \ref{tab:KRbPD}). The proportionality constant accounts for the fraction of the total PD kinetic energy release partitioned into the K photofragment. In the case of Fig. \ref{figPDAnalysis} (b), the $KE$ of ions belonging to the inner and outer Newton spheres are calculated to be 11.67 and 24.52 cm$^{-1}$, respectively.

Fig. \ref{figVMICal}(a) shows a summary of $R$ and $KE$ obtained at different wavelengths. Contributions from the two different PD channels are color-coded. Time-slice analysis was used to extract $R$ for all data points except for the two with the lowest $KE$, where the ion TOF distribution are not sufficiently broad to time-slice. In these two cases, inverse Abel analysis \cite{dribinski2002reconstruction} using the PyAbel software package \cite{hickstein2016pyabel} was employed. The data set is fitted to the function $ R^2 = A^2 KE /V_R $ to determine the calibration constant $A$.

For ``high energy'' calibration, $V_R$ is set to 992 V. We photodissociated KRb into K 4S + Rb 6P and ionized Rb 6P to produce $\textrm{Rb}^+$ ions with $KE$ in the 0 - 450 cm$^{-1}$ range (Fig. \ref{figPDAnalysis} (a)) by varying the UV laser wavelength (photon energy) between 358 nm (27933 cm$^{-1}$) and 340 nm (29412 cm$^{-1}$). The magnetic field is 30G at the KRb cloud location. The ion image acquired at 346 nm and its time-slice are shown in Fig. \ref{figPDAnalysis} (f) and (g). $R$ is determined through inverse Abel analysis for the lowest $KE$ data point and through time-slice analysis for all others. The data analysis follows the same procedures as for the ``low energy'' case and the results are shown in Fig. \ref{figVMICal} (b). All identified Newton spheres are assigned to the K (4S$_{1/2}$) + Rb (6P$_{1/2}$) PD channel. Contributions from the K (4S$_{1/2}$) + Rb (6P$_{3/2}$) channel, which is 77.51 cm$^{-1}$ higher in energy (see Tab. \ref{tab:KRbPD}), were not identified from any of the ion images. The calibration constants for both the ``low energy'' ($15.11 \pm 0.05~\textrm{mm} / \sqrt{\textrm{cm}^{-1}/\textrm{V}}$) and ``high energy'' ($15.60 \pm 0.11~\textrm{mm} / \sqrt{\textrm{cm}^{-1}/\textrm{V}}$) settings obtained from fits are reasonably close to the design value of ($16.44~\textrm{mm} / \sqrt{\textrm{cm}^{-1}/\textrm{V}}$).

\section{Outlook} \label{section: Outlook}

In conclusion, we present the ``ultracold chemistry machine'' and demonstrate its capability to produce ultracold, quantum state-specific reactants, detect all species involved the ensuing reactions, and measure the kinetic energy distribution of particles. The direct detection of both the products and the transient intermediate of the KRb + KRb reaction \cite{hu2019direct} opens the door to further investigations of the reaction dynamics.

With the calibrated ion kinetic energy spectrometer, we aim to study the product state distribution of the KRb + KRb reaction. Using energy conservation, which establishes a unique correspondence between the internal and translational energies of the products, the internal state distribution can be extracted from the velocity-map ion image of the products. In addition,  resonantly-enhanced multi-photon ionization (REMPI) can also provide product quantum state information. Combined use of REMPI and VMI together will provide  the most complete picture of the state distributions of K$_2$ and Rb$_2$.

The transient intermediate complex and its dynamics is another subject of immediate interest for further investigation. In the time-domain, directly measuring the lifetime of the $\textrm{K}_2\textrm{Rb}_2^*$ complex will provide insights into the role of long-lived complexes in bialkali quantum gases, a subject of ongoing theoretical debate \cite{PhysRevLett.123.123402,mayle2013scattering} and experimental investigation \cite{gregory2019sticky,ye2018collisions}. In the frequency-domain, the energy and geometric structures of the complex can be studied through spectroscopy. It may be possible to directly influence the complex using external fields and gain control over the rate or outcome of the reaction.

Beyond studying the reactions between rovibronic ground state molecules, molecule formation via STIRAP also allows reactants in rovibrationally excited states to be prepared. Under ultracold conditions, even small changes in the degree of internal excitation can significantly impact the reaction dynamics, which can manifest in changes of the complex lifetime and the product distribution. Studying reactions starting from different initial states can potentially provide a systematic picture on how the dynamics of complex-forming reactions depend on the density of states and the number of open channels.

Bringing together tools from AMO phyiscs and physical chemistry allows for the exploration of the largely uncharted territory that is ultracold reaction dynamics, with many exciting possibilities and surprises ahead.

\section*{Acknowledgement}
We thank Matthew Nichols and Lingbang Zhu for experimental assistance.
This work was supported from DOE, David and Lucile Packard Foundation, and the NSF through Harvard-MIT CUA. D.D. G. acknowledges supports from the MPHQ.

%%%REFERENCES%%%
\bibliography{ref}
\bibliographystyle{rsc}
\end{document}